\documentclass[11pt,twoside]{article}

\usepackage{asp2014}

\aspSuppressVolSlug
\resetcounters

\begin{document}

\title{Web accessibility trends and implementation in dynamic web applications}

\author{Timothy~W.~Hostetler$^1$,
        Shinyi~Chen,
        Sergi~Blanco-Cuaresma,
        Alberto~Accomazzi,
        Michael~J.~Kurtz,
        Carolyn~S.~Grant,
        Edwin~Henneken,
        Donna~M.~Thompson,
        Roman~Chyla,
        Golnaz~Shapurian,
        Matthew~R.~Templeton,
        Kelly~E.~Lockhart,
        Nemanja~Martinovic,
        Stephen~McDonald,
        Felix~Grezes.
} 
\affil{$^1$Center for Astrophysics | Harvard\ \&\ Smithsonian, Cambridge, MA, USA; \email{timothy.hostetler@cfa.harvard.edu$}}

\paperauthor{Timothy~W.~Hostetler}{thostetler@cfa.harvard.edu}{0000-0001-9238-3667}{Harvard-Smithsonian Center for Astrophysics}{HEAD}{Cambridge}{MA}{02138}{USA}
\paperauthor{Shinyi~Chen}{shinyi.chen@cfa.harvard.edu}{0000-0002-7641-7051}{Harvard-Smithsonian Center for Astrophysics}{HEAD}{Cambridge}{MA}{02138}{USA}
\paperauthor{Sergi~Blanco-Cuaresma}{sblancocuaresma@cfa.harvard.edu}{0000-0002-1584-0171}{Harvard-Smithsonian Center for Astrophysics}{HEAD}{Cambridge}{MA}{02138}{USA}
\paperauthor{Alberto~Accomazzi}{aaccomazzi@cfa.harvard.edu}{0000-0002-4110-3511}{Harvard-Smithsonian Center for Astrophysics}{HEAD}{Cambridge}{MA}{02138}{USA}
\paperauthor{Michael~J.~Kurtz}{kurtz@cfa.harvard.edu}{0000-0002-6949-0090}{Harvard-Smithsonian Center for Astrophysics}{HEAD}{Cambridge}{MA}{02138}{USA}
\paperauthor{Carolyn~S.~Grant}{cgrant@cfa.harvard.edu}{0000-0003-4424-7366}{Harvard-Smithsonian Center for Astrophysics}{HEAD}{Cambridge}{MA}{02138}{USA}
\paperauthor{Edwin~A.~Henneken}{ehenneken@cfa.harvard.edu}{0000-0003-4264-2450}{Harvard-Smithsonian Center for Astrophysics}{HEAD}{Cambridge}{MA}{02138}{USA}
\paperauthor{Donna~M.~Thompson}{dthompson@cfa.harvard.edu}{0000-0001-6870-2365}{Harvard-Smithsonian Center for Astrophysics}{HEAD}{Cambridge}{MA}{02138}{USA}
\paperauthor{Roman~Chyla}{rchyla@cfa.harvard.edu}{0000-0003-3041-2092}{Harvard-Smithsonian Center for Astrophysics}{HEAD}{Cambridge}{MA}{02138}{USA}
\paperauthor{Golnaz~Shapurian}{gshapurian@cfa.harvard.edu}{0000-0001-9759-9811}{Harvard-Smithsonian Center for Astrophysics}{HEAD}{Cambridge}{MA}{02138}{USA}
\paperauthor{Matthew~R.~Templeton}{matthew.templeton@cfa.harvard.edu}{0000-0003-1918-0622}{Harvard-Smithsonian Center for Astrophysics}{HEAD}{Cambridge}{MA}{02138}{USA}
\paperauthor{Kelly~E.~Lockhart}{kelly.lockhart@cfa.harvard.edu}{0000-0002-8130-1440}{Harvard-Smithsonian Center for Astrophysics}{HEAD}{Cambridge}{MA}{02138}{USA}
\paperauthor{Nemanja~Martinovic}{nemanja.martinovic@cfa.harvard.edu}{0000-0002-9485-7296}{Harvard-Smithsonian Center for Astrophysics}{HEAD}{Cambridge}{MA}{02138}{USA}
\paperauthor{Stephen~McDonald}{stephen.mcdonald@cfa.harvard.edu}{0000-0003-1270-0605}{Harvard-Smithsonian Center for Astrophysics}{HEAD}{Cambridge}{MA}{02138}{USA}
\paperauthor{Felix Grezes}{felix.grezes@cfa.harvard.edu}{0000-0001-8714-7774}{Harvard-Smithsonian Center for Astrophysics}{HEAD}{Cambridge}{MA}{02138}{USA}



\begin{abstract}
The NASA Astrophysics Data System (ADS), a critical research service for the astrophysics community, strives to provide the most accessible and inclusive environment for the discovery and exploration of the astronomical literature. Part of this goal involves creating a digital platform that can accommodate everybody, including those with disabilities that would benefit from alternative ways to present the information provided by the website. NASA ADS follows the official Web Content Accessibility Guidelines (WCAG) standard for ensuring accessibility of all its applications, striving to exceed this standard where possible. Through the use of both internal audits and external expert review based on these guidelines, we have identified many areas for improving accessibility in our current web application, and have implemented a number of updates to the UI as a result of this. We present an overview of some current web accessibility trends, discuss our experience incorporating these trends in our web application, and discuss the lessons learned and recommendations for future projects. 
\end{abstract}


\section{Introduction}
Every year the WebAIM\citep{WebAIMSurvey}, a nonprofit organization specializing in accessible web solutions holds a survey of the top 1 million websites using their evaluation tool called WAVE. The organization has found that even in some of the most popular home pages on the web, there were on average 51.4 accessibility errors per page. Many accessibility errors are easily fixed by conforming to web standards, using good web page structure, and following guidelines like those found in the Web Content Accessibility Guidelines\footnote{\url{https://www.w3.org/WAI/standards-guidelines/wcag/}} (WCAG) provided by W3C Web Accessibility Initiative\footnote{\url{https://www.w3.org/WAI/}} (WAI). 
The NASA Astrophysics Data System\footnote{\url{https://ui.adsabs.harvard.edu}} \citep[ADS;][]{2000A&AS..143...41K} web site is a modern single-page application, which presents a unique challenge for web accessibility due to its dynamic content.  Utilizing basic techniques and some special attributes, we can provide accessible content even when the information on the page is changing over time. 

\section{Web Content Accessibility Guidelines}
WCAG provides a set of guidelines that cover a large range of web content and components. The current standard is WCAG 2.1, which is further broken down into three levels of conformance: A (lowest), AA, AAA (highest). There are 4 main sections which are defined: Perceivable, Operable, Understandable and Robust.

\subsection{Perceivable}
Web developers should ensure in their apps, if possible, that there is at least one alternative way for users to get access to content.  Non-text elements like images should utilize \texttt{alt} tags, and videos should have transcripts and/or audio descriptions. HTML standard components like lists and tables should have proper structure and predictability, allowing users utilizing tools to perceive content on the page to have an easier time doing so. In general, any visual changes to the site should not cause a loss of information.

\subsection{Operable}
All content that is accessible by mouse should also be accessible using a keyboard. Content does not induce seizures or physical reactions of any kind and users are given enough time to perform any actions. Make sure the page layout and structure makes finding and getting to content as easy as possible.  All potential input modalities should be taken into account, such as touch and voice.

\subsection{Understandable}
Text is readable and understandable, and presented in a clear and predictable way. Users are helped, where possible, to correct and avoid mistakes. 

\subsection{Robust}
Application markup should be well-structured and reliably interpreted in order to maximise compatibility with existing and future browsers, assistive technologies and any other type of user agent.

\section{Web Accessibility Basics}

\subsection{Page Structure}
Semantic HTML provides structural elements like \texttt{<nav>}, \texttt{<header>}, \texttt{<main>}, and \texttt{<footer>}, these are known as page regions.  Some or all of them should be used to layout a page, since they allow screen readers and other tools to infer some information about the content within. There are also other elements, like \texttt{<section>} and \texttt{<article>} which should be utilized as much as possible, to provide more context to users.

According to a screen reader survey performed by WebAIM, users have relied less and less on page regions to navigate pages.  Instead, nearly \textbf{70\%} of respondents indicated that they prefer to use headings to find information on the page. This is an important insight, since headings (or landmarks) provide a more flexible way of laying out content on the page than the more rigid region tags.  Web apps should therefore be utilizing heading tags to provide "jump" points for screen reader users.

ADS's web app uses heading tags extensively to provide structure and layout. Screen readers will follow the hierarchy of headings to jump to areas of interest or to list the titles of search result, for example.

\subsection{Text Alternatives}
Non-text content like images and icons need to have some kind of textual description. This can be accomplished using semantic tags like \texttt{<figure>} and \texttt{<figcaption>} and in the case of images also using the \texttt{alt} attribute. Some visually hidden inline text can be added next to the icon or image to provide some extra information to screen readers. 

Aria attributes are a set of accessibility-focused attributes which supplement HTML to provide extra information, especially when complex structures are involved or when creating dynamic content.  The use of \texttt{aria} attributes like \texttt{aria-labelledby} can be used to refer to a section title or some other element for example.

\section{Dynamic Content}
One of the challenges with modern web applications is the need for handling page transitions and information change in an accessible way. In ADS, almost every page has some kind of dynamic content which is changing in some way over time. In order to present changing information to screen readers, we can utilize some special attributes. Using the \texttt{role="status"} or \texttt{role="alert"} attribute we can inform screen readers that this element has information in it that will change over time. It will automatically announce to the user when something within this element changes. Another attribute \texttt{aria-atomic="true"} can be provided to have the tool read the entire element text again, not just what changed. This can be quite useful for status messages and error messages, for example.

\section{Tools}
During development on the web, it is useful to regularly be testing that your WCAG level of conformance is being maintained. Some static analysis tools like WAVE and axe DevTools are browser extensions which can be used to directly analyze the website. Screen readers should also be employed as often as possible to get practice with use and to provide some limited real-world experience, without the use of actual users. Some common screen readers are JAWS, NVDA, and VoiceOver.   

\section{Accessibility Audits}
ADS contracted a user experience and accessibility consultancy based at Harvard University called Harvard Web Publishing (HWP). HWP's in-house accessibility experts utilized VoiceOver and NVDA screen readers and provided us with a detailed report of several ADS web pages. Within the provided report, we found many small areas of improvement, which were fixed. 

One benefit to having an external audit like this is to provide us with insight into how a particular user can be confounded by a lack of information or context. By not providing appropriate semantic elements or not labelling form elements properly, screen reader users were unable to understand the purpose of some controls. Also, how we manage focus needed to improve, with some areas inaccessible via keyboard.

\section{Conclusion}
Web accessibility will always be an ongoing effort, with some areas never being able to fully conform to WAI standards. However, all sites should conform at least to the minimal accessibility level (A). At ADS, we continue to improve accessibility in our web application and any future enhancements of our user interface will have the highest conformity as a core goal. 

\acknowledgements The NASA Astrophysics Data System is operated by the Smithsonian Astrophysical Observatory under NASA Cooperative Agreement 80NSSC21M0056.


\bibliography{citations}  

\begin{thebibliography}{}
\expandafter\ifx\csname natexlab\endcsname\relax\def\natexlab#1{#1}\fi
\expandafter\ifx\csname url\endcsname\relax
  \def\url#1{\texttt{#1}}\fi
\expandafter\ifx\csname urlprefix\endcsname\relax\def\urlprefix{URL }\fi
\providecommand{\eprint}[2][]{\url{#2}}

\bibitem[{IDRPP(2021)}]{WebAIMSurvey}
IDRPP 2021, Webaim: Screen reader user survey \#9 results.
  \urlprefix\url{https://webaim.org/projects/screenreadersurvey9}

\bibitem[{{Kurtz} et~al.(2000){Kurtz}, {Eichhorn}, {Accomazzi}, {Grant},
  {Murray}, \& {Watson}}]{2000A&AS..143...41K}
{Kurtz}, M.~J., {Eichhorn}, G., {Accomazzi}, A., {Grant}, C.~S., {Murray},
  S.~S., \& {Watson}, J.~M. 2000, Astronomy and Astrophysics Supplement Series,
  143, 41. \eprint{astro-ph/0002104}

\end{thebibliography}


\end{document}